# Room-Temperature Surface Exciton Polaritons in Colloidal J-Aggregate Flakes

*Carla Estévez-Varela, José Nuno Gama, Miguel Castillo, Adelaide Miranda, Pieter De Beule, Jorge Pérez-Juste, Martin Lopez-Garcia, Isabel Pastoriza-Santos\*, and Sara Núñez-Sánchez\**

C. Estévez-Varela, J. N. Gama, J. Pérez-Juste, I. Pastoriza-Santos

CINBIO, Universidade de Vigo, Campus Universitario As Lagoas, Marcosende, 36310, Vigo, Spain

E-mail: pastoriza@uvigo.es

M. Castillo, A. Miranda, P. De Beule

International Iberian Nanotechnology Laboratory (INL), Avenida Mestre José Veiga s/n 4715-330 Braga, Portugal

M. Lopez-Garcia

Instituto de Óptica, Consejo Superior de Investigaciones Científicas (IO−CSIC), C/Serrano 121, Madrid 28006, Spain

S. Núñez-Sánchez

Instituto de Ciencia de Materiales de Madrid (ICMM), Consejo Superior de Investigaciones Científicas (CSIC), Sor Juana Inés de la Cruz 3, 28049 Madrid, Spain

E-mail: s.nunez.sanchez@csic.es

Funding: Grants: TED2021-130522B-I00, PRE2020-096163 and CNS2023-145364 funded by MCIN/AEI /10.13039/501100011033 and NextGenerationEU/ PRTR; Grant:101129661-ADAPTATION funded by European Union

**Abstract**

J-aggregates are promising organic materials for nanophotonic applications due to their excitonic properties and ability to support surface exciton polaritons at room temperature, providing a robust platform for nanoscale light manipulation. While thin films composed of J-aggregates have demonstrated these advantages, colloidal J-aggregate nanoparticles remain underexplored. Herein, we report the synthesis of colloidal J-aggregate flakes by electrostatic interaction of cyanine molecules (TDBC) and oppositely charged polyelectrolytes (polydiallyldimethylammonium chloride, PDDA). These flakes exhibit colloidal stability maintaining the J-aggregate conformation even in solvents that favoured their monomeric state. The characterization of the colloidal J-aggregate flakes reveals their capability to support surface exciton polaritons at room temperature. This was further confirmed at single-particle level by observing an angular-independent Reststrahlen band near the excitonic resonance. In addition, the colloidal flakes exhibit a strong scattering component that broadens the extinction band and redshifts the photoluminescence, indicating that the colloidal architecture influences the optical response. These findings introduce a versatile colloidal system for constructing excitonic nanostructures tailored for advanced photonic applications.

**1. Introduction**

Surface polaritons are surface electromagnetic waves that propagate along an interface of materials, typically involving at least one medium with negative permittivity, and are coupled to excitations such as plasmons, excitons, or phonons. Among these, surface exciton polaritons (SEPs) arise from the coupling of photons with excitons at the surface of inorganic and organic crystals.[1,2] In inorganic semiconductors, SEPs are usually observed only at cryogenic temperatures due to the low binding energy of Wannier-Mott excitons (~ 0.01 eV), which is lower than the thermal energy at room temperature (0.025 eV).[2] Two-dimensional (2D) semiconductor materials that support SEPs are under extensive study due to their ability to confine charge and heat transport within a plane.[3] Examples of inorganic semiconductors currently explored for SEP applications include graphene, transition metal dichalcogenides (TMDCs), hexagonal boron nitride (h-BN), or phosphorene.[4] Ideally, a suitable semiconductor should exhibit a suitable bandgap and excellent carrier transport properties with decreasing body thickness, along with low defect density, long-term stability, scalability, ease of integration and compatibility with Si technology.[5] Summarising, these materials offer remarkable carrier mobility, high surface-to-volume ratios and enhanced surface activity.[4–6]

In contrast to inorganic systems, organic materials offer a significant advantage for SEP applications: their excitons are of the Frenkel type, with binding energies on the order of 1 eV, substantially higher than the thermal energy at room temperature. This enables the observation of SEPs in organic crystals at room temperature.[2,7] A representative example is the case of the J-aggregate materials, supramolecular assemblies exhibiting a narrowband optical response due to the delocalised Frenkel excitons. Owing to this narrowband behaviour, J-aggregates can exhibit a Reststrahlen band with negative values of the real part of the dielectric function at specific wavelengths, even though they lack free charge carriers.[8,9] This renders them suitable for supporting SEPs at visible wavelengths through purely excitonic mechanisms. Previous studies demonstrated the SEP propagation in thin films composed of cyanine J-aggregates, offering a viable platform to manipulate light at the nanoscale using organic materials.[1,2,9,10] Nevertheless, to the best of our knowledge, there are no reports of SEP propagation in colloidal J-aggregate nanoparticles.

The optical response of J-aggregates can be modelled by Lorentzian oscillators, which allow to support open cavity SEP at the interface, similar to surface plasmon polaritons in metal thin films.[11,12] However, the optical properties depend on the molecular arrangement, and thus, their stability is highly sensitive to environmental conditions. Variations in pH, ionic strength, or solvent polarity can disrupt the supramolecular structure, breaking J-aggregates into monomers and losing their characteristic optical properties. Different stabilization strategies have been explored, *e.g.,* through polymer[13] or silica encapsulation,[14] to preserve their properties and expand the applicability of J-aggregates in photonics[15] or bio-imaging.[16,17]

In this work, we present a strategy to fabricate colloidal organic flakes composed of J-aggregates that support SEPs at room temperature, exhibiting enhanced stability under different solvent conditions. The method relies on electrostatic interactions between J-aggregates of 5,6-Dichloro-2-[[5,6-dichloro-1-ethyl-3-(4-sulfobutyl)-benzimidazol-2-ylidene]-propenyl]-1-ethyl-3-(4-sulfobutyl)-benzimidazolium hydroxide (TDBC) and a cationic polyelectrolyte, polydiallyldimethylammonium chloride (PDDA). The optical properties of the organic flakes are characterized by UV-vis and fluorescence spectroscopy. Their morphology and surface topology are elucidated through Transmission Electron Microscopy (TEM), Atomic Force Microscopy (AFM), and Fluorescence Optical Microscopy. Finally, angular-independent, narrowband reflectivity features—indicative of a Reststrahlen band and negative permittivity—are analyzed using visible Fourier imaging spectroscopy (Vis-FIS). Vis-FIS further confirms incident light-SEP coupling in the visible range, establishing these colloidal flakes as viable platforms for room-temperature exciton-polariton applications.

## 2. Results and discussion

### 2.1. Formation of colloidal organic flakes based on J-aggregates

Taking advantage of the fact that TDBC is a cyanine dye with negatively charged sulphonate groups, colloidal organic flakes were synthesized via electrostatic interactions between preformed TDBC J-aggregates with the cationic polyelectrolyte PDDA in the presence of 0.5 M NaCl (**Figure 1A**). The addition of NaCl favoured the formation of TDBC J-aggregates by screening electrostatic repulsions among dye molecules. The effect of the J-aggregate to PDDA molar ratio on the formation of colloidal nanostructures was systematically studied by keeping constant the final PDDA concentration at 5.6 µM and varying the TDBC J-aggregates to obtain molar ratios from 3 to 150 (see the methods section for details). Zeta potential analysis of the resulting structures revealed that the negative charge of the TDBC J-aggregates (-43.6 ± 3.2 mV) shifted to positive values upon interaction with PDDA (**Figure S1,** Supporting Information). Notably, Zeta potential values of around +50 mV were recorded across all the J-aggregate:PDDA molar ratios, suggesting the formation of highly stable colloidal structures in water. The visible extinction spectra of the colloidal dispersions further validated the formation and stability of the J-aggregate-based structures (**Figure** 1B and **S2,** Supporting Information (black curves)). All samples exhibited a sharp absorption band at 587 nm, characteristic of TDBC J-aggregates, with no evidence of TDBC monomers (no band at 520 nm). This indicates that PDDA stabilizes the J-aggregate state of TDBC in water, irrespective of J-aggregate:PDDA molar ratio. It is noteworthy that in water and the absence of PDDA, the TDBC exists as a mixture of J-aggregates and monomers.

To test the stability of the colloidal structures under unfavourable conditions, samples were redispersed in an EtOH:$H_2O$ mixture of 1:1, which typically favours the dissociation of J-aggregates. As shown in Figure 1B and S2, Supporting Information (red curves), the pristine TDBC completely dissociates into monomers, exhibiting a dominant band at 520 nm. However, the colloidal structures retained their aggregated state and spectral signature at 587 nm when the J-aggregate:PDDA molar ratio is higher than 15, demonstrating stability even in destabilizing solvents. At molar ratios below 3, both J-aggregate and monomer bands were observed, indicating the lack of stability of the J-aggregate in these colloidal organic flakes. Additionally, colloidal structures with molar ratios above 30 maintained their stability even in pure ethanol or methanol, solvents that strongly favour the monomeric TDBC state (Figure 1C).

The morphology of these organic structures was examined using TEM (**Figure** 1D and **S3**, Supporting Information). TEM images revealed a predominantly flake-like nanostructure with well-defined edges and considerable polydispersity, with sizes ranging from 100 nm to several micrometers for J-aggregate:PDDA molar ratios spanning from 15 to 150. At a molar ratio of 3, while a small amount of flakes was observed, other features with non-defined morphology were present (Figure S3, Supporting Information). These results align with their corresponding extinction spectrum (Figure 1B), which indicates the coexistence of monomers and J-aggregates.

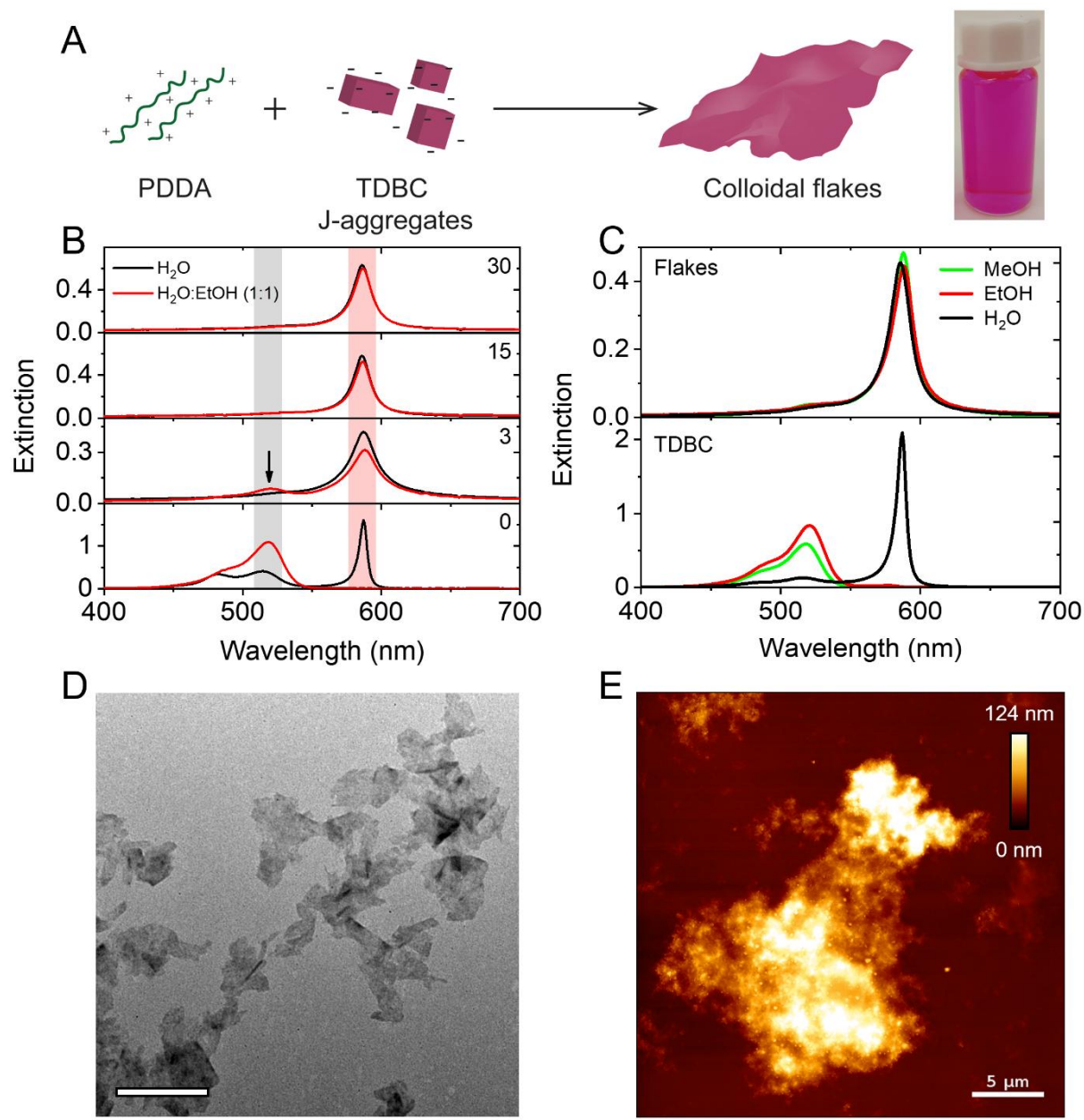

**Figure 1**. A) Schematic representation of the synthesis of the colloidal flakes. B) Stability test: extinction spectra of colloidal flakes prepared with different J-aggregate:PDDA molar ratios (3, 15 and 30) in water (black) and in a 1:1 $H_2O$:EtOH mixture (red). The grey and red shades indicate the position of the absorption band of TDBC monomer and J-aggregate, respectively.

The spectrum for pristine TDBC in H$_2$O (black) and a 1:1 H$_2$O:EtOH mixture is included as 0 for comparison. C) Extinction spectra of colloidal flakes (molar ratio 30) (top) and pristine TDBC dye (bottom) in methanol (green), ethanol (red), and water (black). D) Representative TEM image of organic flakes obtained with a J-aggregate:PDDA molar ratio of 30. The scale bar is 1 μm. E) Representative topographic AFM image of a single flake obtained with J-aggregate:PDDA molar ratio of 30.

Colloidal J-aggregate flakes were immobilised onto cover glass substrates via electrostatic deposition for AFM and fluorescence microscopy measurements (see experimental methods for details). The optical and luminescence imaging of the deposited flakes further confirmed the broad size distribution of the samples (**Figure S4**, Supporting Information). The flake-like morphology was additionally corroborated by AFM, as shown in **Figure** 1E and **S5,** Supporting Information. Interestingly, AFM enabled estimation of the thickness of the flake-like structures, which ranged from approximately 10 to 125 nm. Thicker flakes were generally associated with larger lateral dimensions. Interestingly, confocal fluorescence imaging revealed that the relative abundance of larger flakes increased with higher J-aggregate:PDDA molar ratios (Figure S4**,** Supporting Information). While no significant morphological differences were observed among samples with ratios above 15, fluorescence data suggest that larger flakes contribute more prominently to the overall emission signal as the molar ratio increases, likely due to their greater size.

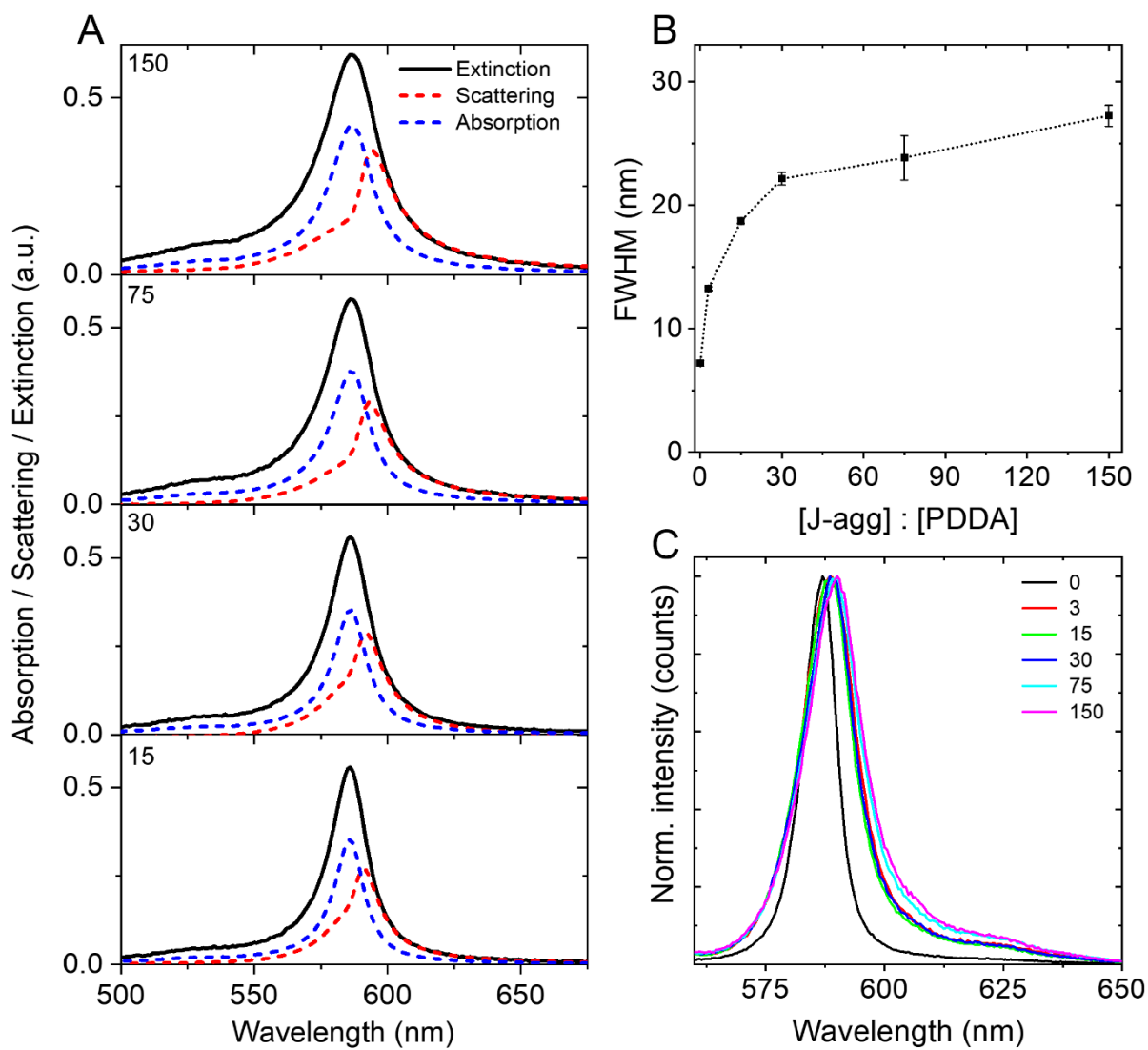

**Figure 2.** A) Extinction, scattering and absorption spectra of colloidal J-aggregate flakes prepared with different J-aggregate:PDDA molar ratio: 15, 30, 75, and 150, as indicated. The black solid lines represent extinction, red dashed lines represent scattering, and blue dashed lines represent absorption. B) Corresponding Full Width at Half Maximum (FWHM) of the extinction peak as a function of J-aggregate:PDDA molar ratios. Error bars represent three different colloidal synteses, except for the 75 molar ratio, which was derivedfrom two experiments. C) Corresponding normalized photoluminescence spectra of colloidal flakes in $H_2O$ excited at 550 nm. The spectrum for pristine TDBC in $H_2O$ is included as 0 for comparison.

A more detailed analysis of the optical response of flakes in water (**Figure 2A**), prepared at molar ratios above 15, revealed that their main extinction band remained centred at the characteristic J-aggregate transition (~587 nm) but exhibiting noticeable broadening compared to pristine TDBC J-aggregates. Specifically, the full width at half maximum (FWHM) of the extinction band increased with molar ratio (Figure 2B). To further investigate this behaviour, we performed integrating sphere analysis to decouple the scattering and absorption components from the extinction (see methods). The results revealed distinct spectral profiles for absorption and scattering, with the absorption peak consistently located at shorter wavelengths than the scattering one. Moreover, the absorption spectra displayed a symmetric peak centered at 587 nm regardless of the molar ratio, aligning with the extinction peak of J-aggregates in aqueous solution. This feature can be attributed to the intrinsic molecular absorption of the J-aggregates forming the flakes.

In contrast, the scattering response varied with the J-aggregate:PDDA molar ratio, suggesting a dependence on the shape and size of the colloidal flakes. Interestingly, the scattering spectra exhibited a truncated profile at the absorption peak of the J-aggregates, with a scattering peak more intense at longer wavelengths. This redshift in the scattering peak with increasing molar ratio can be associated with a higher presence of larger colloidal flakes, which are more efficient scatterers, particularly in regions of high refractive index contrast (i.e., wavelengths longer than the excitonic transition). These findings suggest that the observed broadening of the extinction response with the J-aggregate:PDDA molar ratios can be associated with a shift to longer wavelengths of the scattering component due to the increase of the size of the J-aggregate flakes while the intrinsic absorption of the J-aggregates composing the colloids remains stable under different preparation conditions.

Additionally, photoluminescence (PL) measurements exciting at 550 nm reveal a small Stokes shift of around 2 – 4 nm depending on the J-aggregate:PDDA molar ratio, consistent with the narrow excitonic nature of J-aggregates (Figure 2C). However, a progressive redshift of the PL maximum was observed with increasing J-aggregate:PDDA molar ratio. Since the absorption peak and spectrum remain unchanged (Figure 2A) with the J-aggregate:PDDA molar ratios (Figure 2B), this PL shift cannot be attributed to intrinsic changes in the excitonic transitions of the constituent J-aggregates. Instead, it is more likely due to modifications in the photonic environment surrounding the emitters influenced by the scattering effects. As demonstrated by the integrating sphere measurements (Figure 2A), increasing the molar ratios leads to a redshift in the scattering component relative to absorption. These scattering contributions may alter the local density of optical states, thereby affecting the emission spectrum and shifting it to longer wavelengths. A plausible explanation is that the scattering promotes radiative recombination at longer wavelengths through a mechanism analogous to the Purcell effect. In this context, the observed PL redshift does not arise from intrinsic changes in the excitonic properties of the J-aggregates, but rather from light–matter interactions among J-aggregates, modulated by the response of the colloids. To validate this interpretation, further time-resolved PL experiments will be essential, particularly to assess whether increased scattering correlates with a shortening in emission lifetime, thereby supporting the hypothesis of enhanced radiative decay mediated by the photonic environment.

**2.2. Single-flake optical characterization.**

The optical response of the individual J-aggregate colloidal flakes was investigated using angle-resolved reflectance spectroscopy, performed via Fourier imaging spectrometry (FIS) in two configurations: i) air configuration (**Figure S6A**, Supporting Information) for single flake reflectance measurements and ii) Kretschman Prism Coupling configuration for surface exciton polariton coupling in single flakes (Figure S6B, Supporting Information).[18] The use of a high numerical aperture and high magnification objective in the FIS setup enabled precise focusing on small regions, allowing for the study of individual flake analysis of their reflectance properties as a function of the incidence angle. For this study, flakes obtained with a J-aggregate:PDDA molar ratio of 30 were selected. Due to the inherent size polydispersity of the sample, the flakes were categorized into three groups for analysis: small (<15 μm), medium (≈15 μm), and large (>15 μm) for their optical analysis. The angle-resolved reflectance spectrum of a representative medium-sized flake in air displays a broad, high-intensity reflectance band that remains invariant respect to the angle of incidence (**Figure 3A**). This angular independence confirms that the band originates from intrinsic material properties rather than photonic effects. Notably, the reflectance band is consistently blue-shifted relative to the J-aggregate excitonic transition, which is characteristic of resonant bulk materials showing a Reststrahlen band, where the real part of the permittivity achieves negative values. The presence of a strong reflectance peak matching the spectral region of the TDBC Reststrahlen band [9] (**Figure S7,** Supporting Information) suggests that the colloidal flakes achieve dielectric function values comparable to those of J-aggregate thin films. These findings indicate that the required conditions to support SEPs, driven purely by excitonic resonances and without the involvement of free carriers, are achieved.

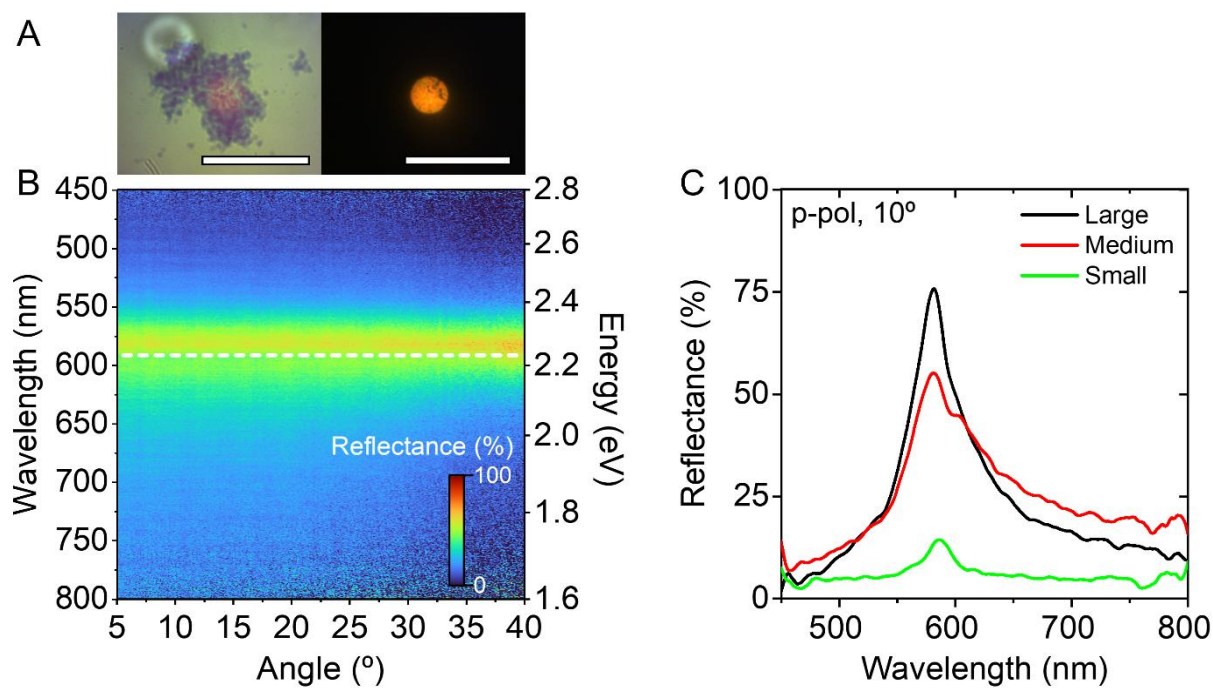

**Figure 3**. A) White-LED transmission (left panel) and halogen-lamp epi-illumination (right panel) microscopy images of the analysed flake. Note that all the spectroscopic characterization was performed using a halogen lamp in epi-illumination configuration. B) Single-flake reflectance as a function of the incident angle for a medium-sized organic flake (ca. 15 μm in diameter) obtained with a J-aggregate:PDDA molar ratio of 30. The white dotted line on top of the contour plot indicates the wavelength of the J-aggregate absorption band at 587 nm. The scale bar is 20 μm. C) Reflectance spectra at 10º for p-polarized incident light for small (<15 μm), medium (≈15 μm), and large (>15 μm).

The comparison of these results with those obtained with small- and large-sized flakes (**Figure S8**, Supporting Information) revealed a systematic increase of the reflectance band intensity with the flake size (Figure 3C). The center of this contrast is also at lower wavelengths than the J-aggregate absorption (white dotted line in Figure 3A), matching the minimum of the oscillator of the dielectric function.[8,9] This supports the interpretation of the observed feature as a bulk optical property arising from the negative permittivity region of the material. Similar reflectance behaviour was found for flakes of the same size prepared with different J-aggregate:PDDA molar ratios, suggesting that the optical response is governed by the flake dimensions (particularly thickness) rather than the specific preparation conditions (**Figure S9**, Supporting information). This conclusion is supported by AFM measurements, which showed a correlation between flake size and thickness with larger flakes generally being thicker (Figure

S5). This morphological trend is consistent with the increased optical contrast observed in FIS measurements.

Next, we analyzed the excitation of SEP at single particle level using FIS in the configuration of attenuated total reflection (ATR) (Figure S6B, Supporting Information).[10] **Figure 4A-B** shows p- and s-polarized angular reflectance for a medium-sized colloidal flake prepared at a J-aggregate:PDDA molar ratio of 30 (**Figure S10**, Supporting information). For p-polarized light, a distinct angle-dependent dip in reflectance is observed at shorter wavelengths than the excitonic absorption of the colloidal flakes for incident angles exceeding the critical angle of the glass substrate (41.2°). This dip is assigned to the coupling of the evanescent electric field, generated by the total internal reflection at the glass interface, with the SEP at the flake-air interface.[19] This coupling observed only in the wavelength range where the material exhibits a Reststrahlen band, enabling the material to achieve negative values of the real part of its dielectric function. To further investigate the angular coupling conditions, fixed-wavelength plots were extracted at 410, 570, and 725 nm for both p-and s-polarized light (Figure 4C-D). At 410 and 725 nm, the flake behaves as a non-resonant dielectric with optical properties like those of the glass substrate, displaying the expected TIR profile: near-zero reflectance below the critical angle, which increases sharply to nearly 100% beyond it. This behavior is consistent for both p- and s-polarizations, confirming the absence of surface mode coupling in these spectral regions. However, at 570 nm under p-polarization, a pronounced minimum in reflectance appears at angles beyond the critical angle, indicating resonant coupling to the SEP. Furthermore, at the critical angle, the sharp increase in reflectance under p-polarization can be interpreted as a Fano-type resonance, resulting from the anti-crossing between the SEP and the light line, a signature of strong light–matter interaction in surface-confined systems.[19] This phenomenon has been previously reported for J-aggregate-based thin films prepared by spin coating [1,2,10,20] and is demonstrated here for the first time in colloidal J-aggregate flakes.

For s-polarization, at wavelengths shorter than the colloidal flakes absorption peak, an angle-independent attenuated reflection is observed beyond the light line dominated by material absorption and a near-zero refractive index region.[20] At longer wavelengths, the refractive index increases, and the optical response mirrors that of a cover glass, exhibiting the reverse trend.

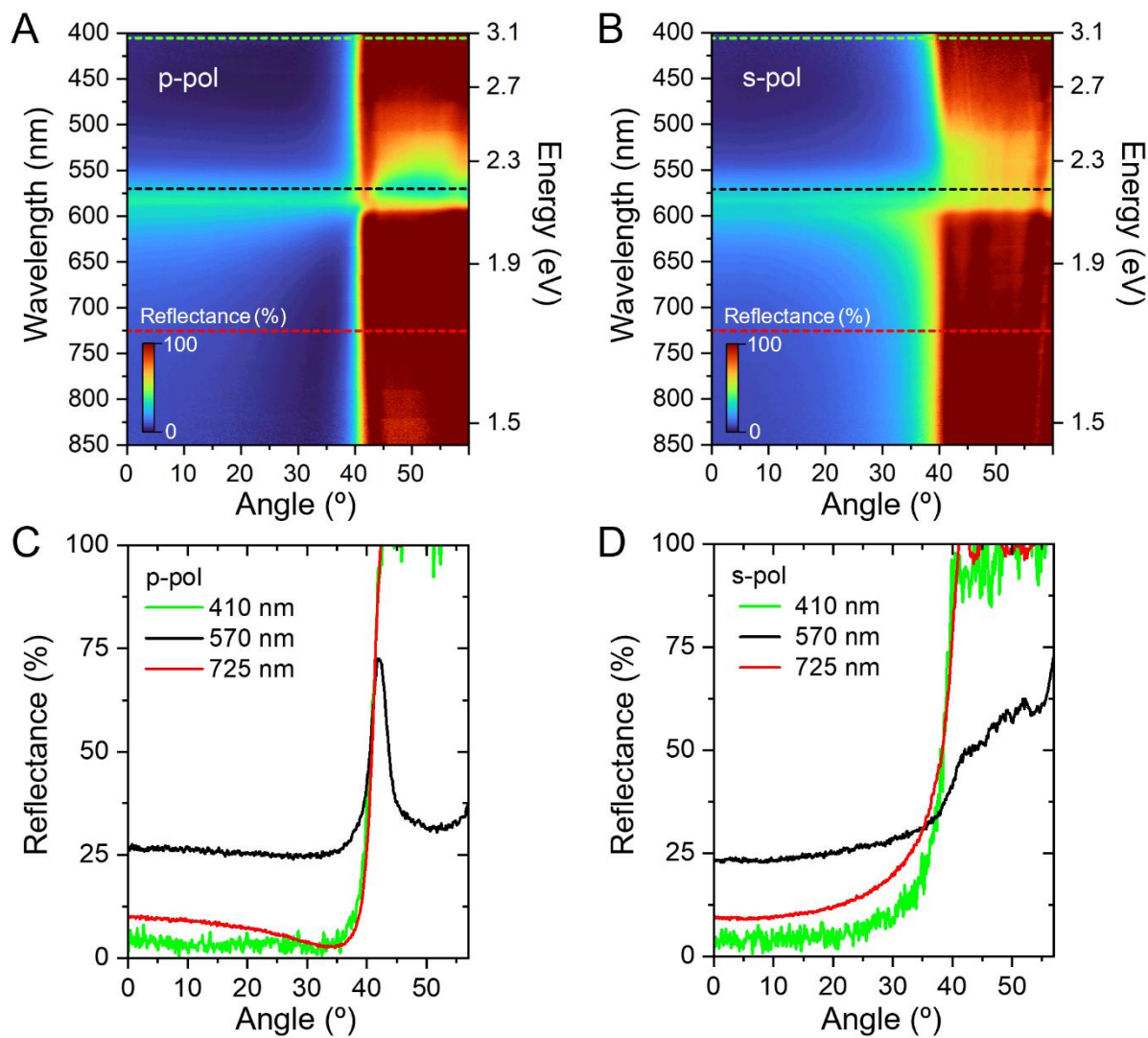

**Figure 4.** A-B) Single-flake p-polarised (A) and s-polarised (B) reflectance as a function of the incident angle in a Kretschmann prism coupling configuration for an organic flake with J-aggregate to polymer molar ratio of 30. C-D) Reflectance spectrum at 410, 570 and 725 nm, as a function of the incident angle for p-polarised (C) and s-polarised (D) light. Note: Reflectance values exceeding 100% are due to the use of a silver mirror as reference, which has lower reflectance than the total internal reflection condition of the prism.

Overall, these results demonstrate that colloidal TDBC-based flakes can support SEPs at room temperature under ATR conditions, despite their non-continuous and non-uniform morphology. The angular and spectral behavior of the reflectance dip confirms the presence of a dispersive surface mode, tightly confined at the flake–air interface and dependent on the material's excitonic resonance. This provides a new pathway to study light–matter coupling at room temperature using solution-processable, soft-matter nanostructures, without requiring nanofabrication or epitaxial growth as in traditional organic thin-film systems.

## 3. Conclusions

In this work, we have developed a method to obtain colloidal J-aggregate flakes via electrostatic interactions between TDBC and PDDA. Interestingly, these organic flakes not only retain the optical properties of the J-aggregates, absorption and PL, but also exhibiting enhanced colloidal stability, even in solvents that typically favor the monomeric state. Moreover, despite the size polydispersity, the flakes exhibit a strong scattering component, originating from their high refractive index and the formation of a Reststrahlen band near the excitonic resonance. This scattering significantly influences the response of the colloidal flakes leading to a broadening of the extinction band with respect to the J-aggregates in solution and a redshift of the photoluminescence response due to changes in the local photonic environment.

Single-flake optical analysis further reveals an angular-independent narrowband reflectivity associated with the Reststrahlen band near the excitonic resonance. Additionally, within this spectral region, the colloidal flakes support SEPs at room temperature at wavelengths shorter than the excitonic peak, resembling the behaviour of 2D-semiconductor materials, which typically require cryogenic conditions to observe such effects.

These results demonstrate the dual potential for exploiting light–matter interactions in colloidal J-aggregate systems: from broadband scattering effects to resonant polariton excitation. This work introduces a scalable, solution-processable strategy for harnessing organic SEPs at room temperature, with potential applications in photonic, optoelectronic, and quantum technologies. Furthermore, this methodology can be extended to other J-aggregate species, opening a route to stabilise J-aggregates within photonic structures across the visible to telecom range via organic chemical design.

## 4. Experimental Section/Methods

*Materials:* Poly(diallyldimethylammonium chloride) (PDDA, Mw: 100,000-200,000, 20 wt. % in $H_2O$), sodium chloride (NaCl, 99%), and methanol (MeOH, 99.9%) were purchased from Sigma Aldrich. 5,6-dichloro-2-[[5,6-dichloro-1-ethyl-3-(4-sulfobutyl)-benzimidazol-2-ylidene]-propenyl]-1-ethyl-3-(4-sulfobutyl)-benzimidazolium hydroxide sodium salt (TDBC) was provided by Few Chemicals. Absolute grade ethanol and Milli-Q water (18.3 MΩ·cm) were used as solvents.

*Synthesis of colloidal flakes:* Colloidal J-aggregate flakes were obtained through an electrostatic mediated methodology. 1 mL of an aqueous solution of TDBC (concentrations ranging from 0.1 mM to 5 mM) was added quickly under bath sonication to 5 mL of an aqueous solution containing 7 μM PDDA and 0.5 M NaCl. After sonication for 15 min, the resulting flakes were centrifuged (9000 rpm, 15 min) and the pellet was redispersed in 6 mL of 0.17 M NaCl in water. This process was repeated twice. Finally, the flakes were redispersed in 5 mL of water.

*Deposition of colloidal flakes on cover slides.* The colloidal J-aggregate flakes were deposited on cover glass slides (0.17 mm-thick microscopy standard) by immersion in the colloidal flake dispersion. First, the cover slide is cleaned in a piranha solution (mixture 3:1 of $H_2SO_4:H_2O_2$) for 20 min and then rinsed in Milli-Q water for 10 min twice. After that, the cover slide is dried by compressed air and immersed in a colloidal flake dispersion for 10 min. After that, the cover slide is dried at room temperature on top a filter paper. Since both sides of the cover slide are functionalised, one of the sides is cleaned with ethanol before measurements.

*Integrating sphere measurements.* Optical measurements were performed using an integrating sphere equipped with two output ports: a transmission port and a diffuse reflectance port positioned at 90° relative to the incident beam. During measurements, the cuvette with the dispersions of colloidal J-aggregate flakes was located at the centre of the integrating sphere. A cuvette with the solvent (water) was used for normalisation purposes. The extinction spectrum was obtained from the direct transmission measurement, with the reflectance port blocked by a diffuse reflector. The scattering component was measured from the diffuse reflectance port, using a light absorber to block the transmitted beam. The absorption was then calculated from the relation with the extinction and scattering components. More details are explained in detail in previous work.[21]

*Optical characterization of single flakes.* The angle and polarization resolved reflectance of single flakes was obtained using two different configurations of the FIS setup. A halogen lamp covering the UV-Vis-NIR spectral range was focused down to a controllable size spot that allows the selection of single areas over the flakes. The reflectance of the flake was then collected by the same high NA lens in an epi-illumination configuration and the back focal plane image of the lens was projected into a UV-Vis spectrograph (Princeton Instruments, Acton SpectraPro SP-2150 with attached CCD camera QImaging Retiga R6 USB3.0 Colour). In the case of measurements in air (Figure S6a) we used a Nikon Flour 40X NA=0.75 objective lens, which allows inspection of the angular reflectance between -48- and 48-degrees angles of incidence. In case of SEP propagation, it can only occur by inducing an evanescent electric field at the flake/air interface. Therefore, light was coupled at the flake surface using a Kretschmann prism-coupling configuration (Figure S6B, Supporting Information), which ensures coupling with wavevectors beyond the glass/air light line in k-space. Given the size of the flakes, we used a high NA oil immersion lens (Nikon 100X Plan Apo NA=1.45) which collects reflected light with angles between -72.78º and 72.78. Due to the small working distance of this lens, we use 0.17 mm-thick microscopy standard cover slide as a substrate for the deposition of the colloidal flakes. The mirror used as reference for normalization was a silver-coated -P01 >97% for 450 nm - 2 μm from THORLABS, and the final normalized data was adjusted according to the reference reflectance data provided for the product at an angle of incidence (AOI) of 45º.

*Characterization:* UV-vis absorption spectra were obtained using an Agilent 8453 UV-vis spectrophotometer with glass and quartz cuvettes of 1 mm optical path. Fluorescence measurements were obtained with a FluoroMax-3 spectrofluorometer (HORIBA Scientific, Japan). TEM images were collected with a JEOL JEM 1010 operating at 100 kV. For the TEM grids, 3-4 drops of 20 μL were cast on a carbon-coated copper grid over filter paper and dried in air at room temperature before the observation. Zeta potential measurements were obtained using a Zetasizer nano ZS, model ZEN 3600 and cuvettes model DTS 1070. Nikon NiE Fluorescence Direct Microscope with image analysis (20X objective and TRITC filter) was used to collect the optical microscope images. Atomic force microscopy (JPK Nanowizard 3 AFM, Bruker Nano GmbH, Germany) topographic measurements in quantitative imaging mode were obtained using a PPP-CONTR (NanoSensors) out-of-the-box cantilevers. The cantilevers with a nominal spring constant of 13 KHz have an aluminum coating layer on the detector side and a radius of curvature of less than 7 nm. AFM images were acquired with a set point of 1.3 nN and a z-length of 2070 nm. Image analysis was performed using the JPK Data Processing software (version 6.1.120). Visible Fourier imaging spectroscopy was performed with a Krestchmann prism-coupling configuration of an objective N*ikon Plan Apo Lambda 100x 1.45 Oil CFI Eclipse ∞/0.17*. Microscope images of section 2.2 were obtained by a *THORLabs 8051C-USB - 8 MP Color* CCD camera. Fourier imaging spectrometry was collected *by QImaging Retiga R6 USB3.0 Colour* CCD camera from an *Acton SpectraPro SP-1250* spectrograph.

**Supporting Information**

Supporting Information is available from the Wiley Online Library.

**Conflict of interest**

The authors declare no conflicts of interest

**Data Availability Statement**

The data that support the findings of this study are available from the corresponding author upon reasonable request.

**Acknowledgements**

This work was supported by Grant TED2021-130522B-I00 funded by MCIN/AEI /10.13039/501100011033 and NextGenerationEU/ PRTR - BDNS 598843 - CATARSIS and the European Union under Grant Agreement #101129661-ADAPTATION. M.L.-G. thanks the Grant CNS2023-145364 funded by MICIU/AEI/10.13039/501100011033 and by "European Union NextGenerationEU/PRTR". C.E.-V. acknowledges an FPI Fellowship from MCIN/AEI/10.13039/501100011033 (Grant No. PRE2020-096163).

Supporting Information

# Room-Temperature Surface Exciton Polaritons in Colloidal J-Aggregate Flakes

*Carla Estévez-Varela, José Nuno Gama, Miguel Castillo, Adelaide Miranda, Pieter De Beule, Jorge Pérez-Juste, Martin Lopez-Garcia, Isabel Pastoriza-Santos\*, and Sara Núñez-Sánchez\**

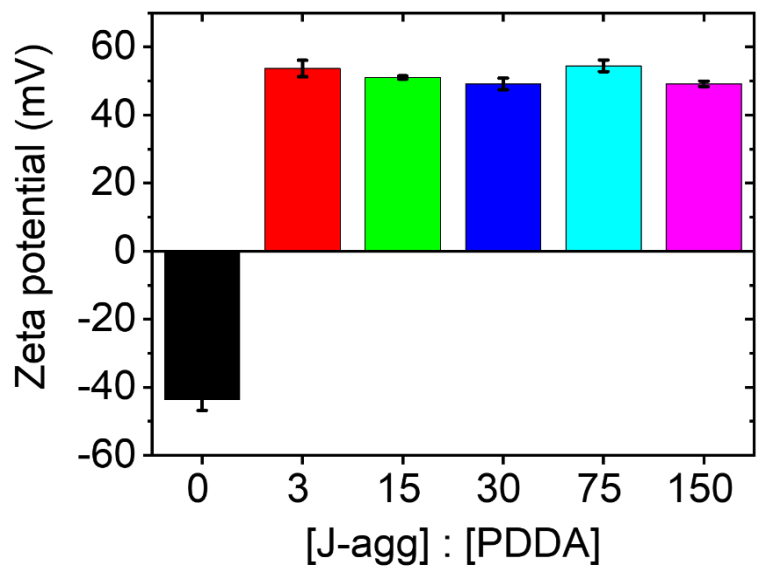

**Figure S1**. Zeta potential of colloidal flakes prepared with different J-aggregate:PDDA molar ratios: 0 (only J-aggregate, black), 3 (red), 15 (green), 30 (blue), 75 (cyan), and 150 (magenta).

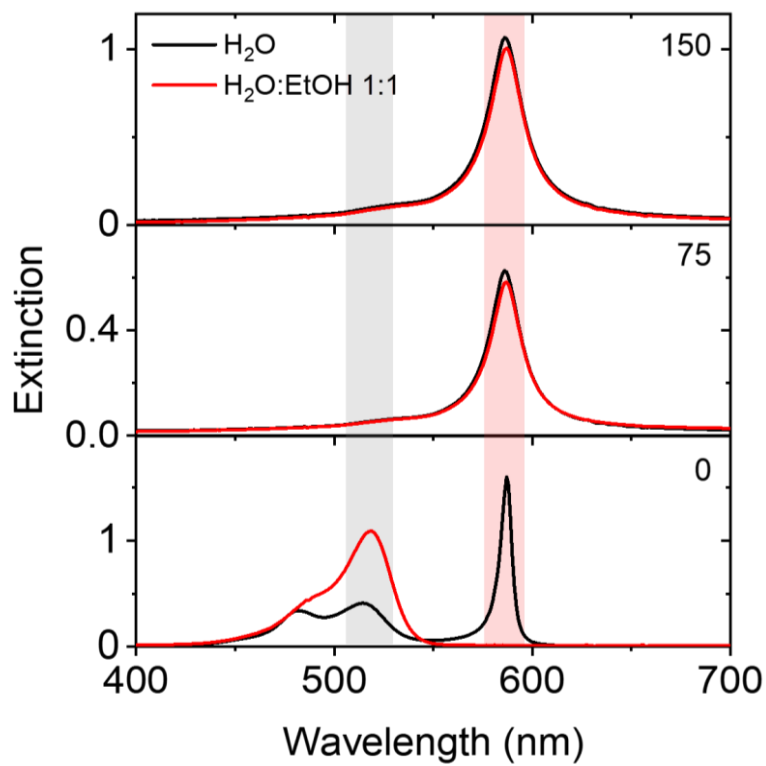

**Figure S2**. Extinction spectra of colloidal flakes prepared with different J-aggregate:PDDA molar ratio of 0, 75 and 150, as indicated, in $H_2O$ (black) and 1:1 $H_2O$:EtOH mixture (red). The spectrum for pristine TBDC dye in $H_2O$ (black) and 1:1 $H_2O$:EtOH mixture is included as 0 for comparison.

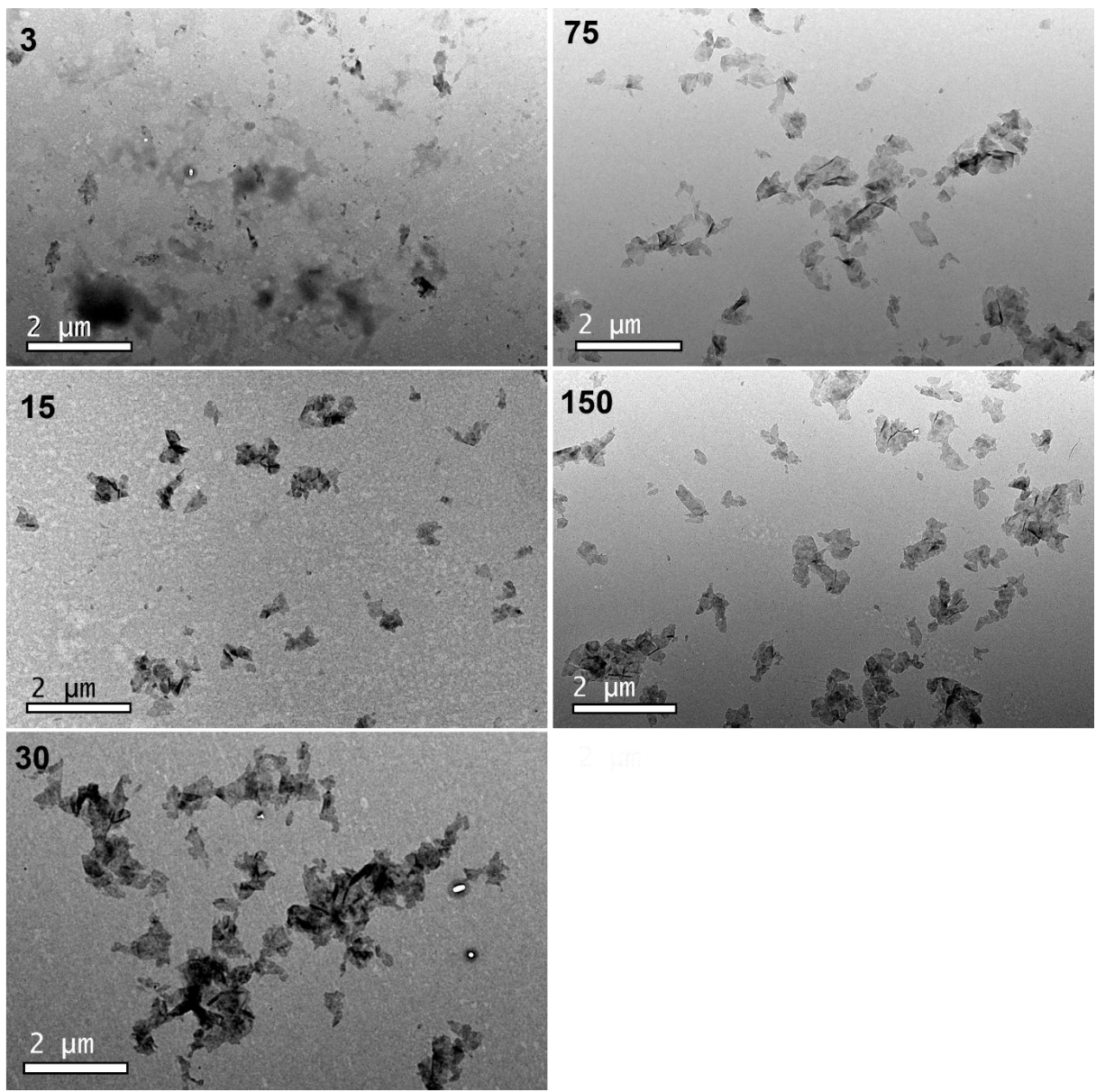

**Figure S3**. Transmission Electron Microscopy images of the colloidal flakes in $H_2O$ obtained with J-aggregate:PDDA molar ratio from 3 to 150, as indicated.

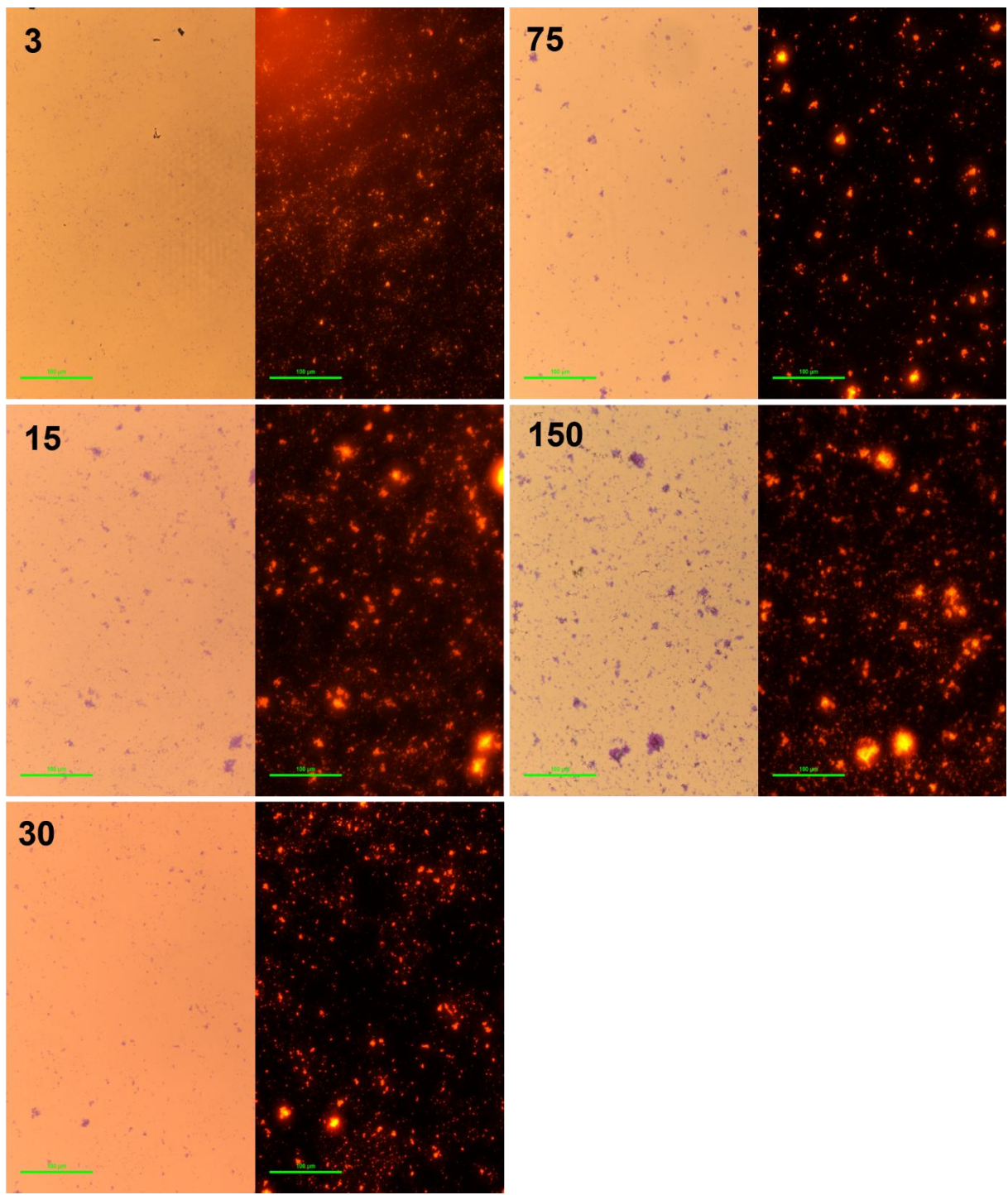

**Figure S4**. Optical microscope images merged with fluorescence of the colloidal flakes prepared with different J-aggregate:PDDA molar ratios from 3 to 150, as indicated. The scale bar is 100 μM.

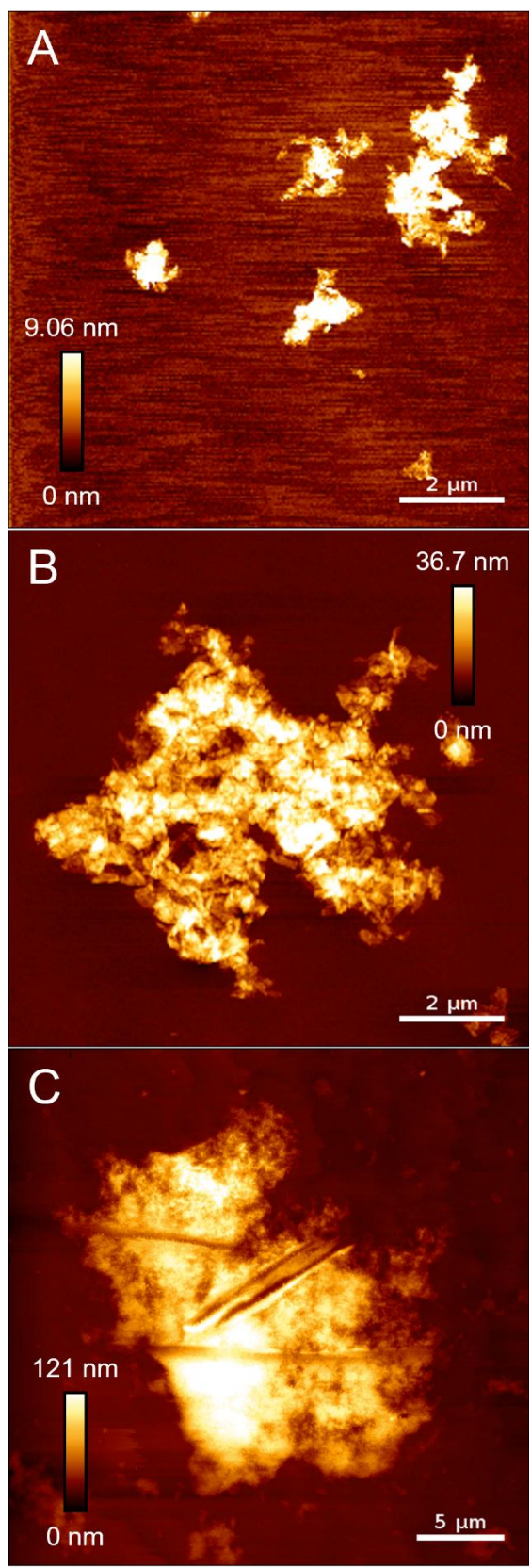

**Figure S5**. Representative topographic AFM images of single flakes of various thicknesses obtained with a J-aggregate:PDDA molar ratio of 30.

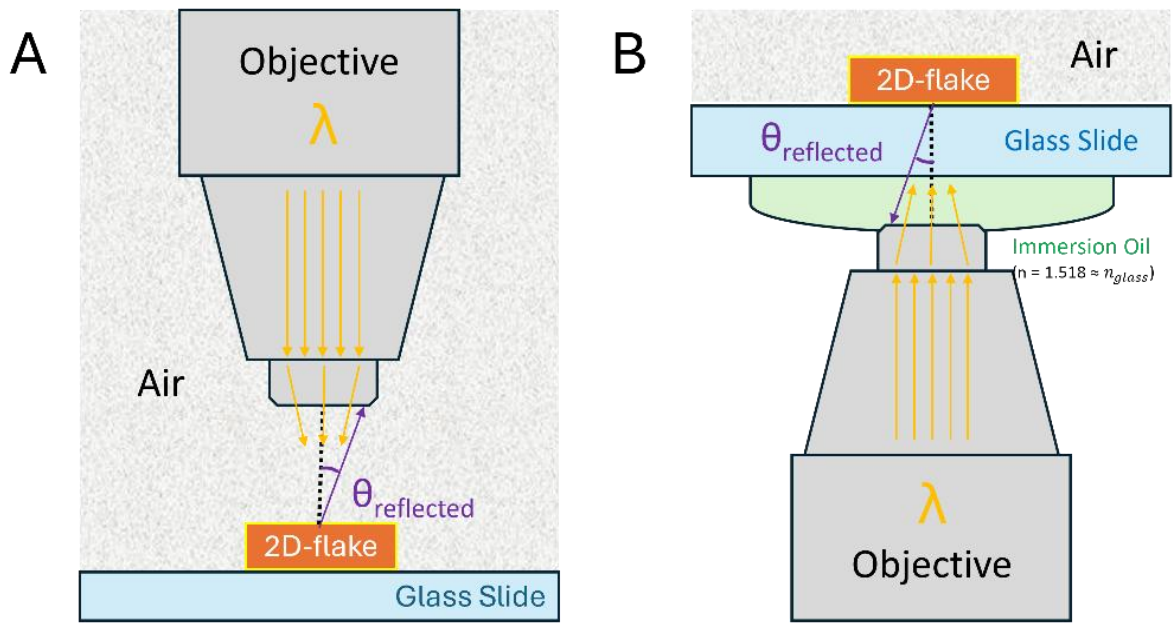

**Figure S6**. Experimental diagram of the Fourier Imaging Spectroscopy setup used to obtain single particle angular reflectance. A) Air objective configuration. B) Kretschman Prism Coupling configuration.

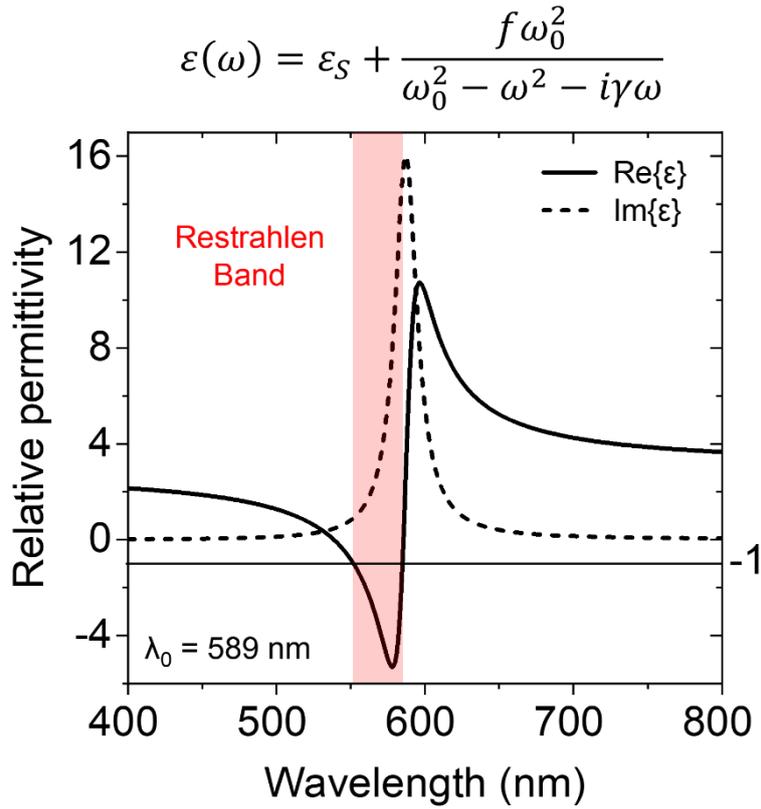

**Figure S7**. Example of a Lorentz oscillator model for permittivity. Analytical equation for the model is on top. Used parameters were $\varepsilon_s$=2.58, baseline, $f$=0.5, oscillator strength, $\omega=\frac{2\pi}{\lambda}$, angular frequency, $\omega_0=\frac{2\pi}{\lambda_0}$, oscillator resonance, $\gamma$=1, damping factor following reference [1]. Red shaded area corresponds to the Reststrahlen band.

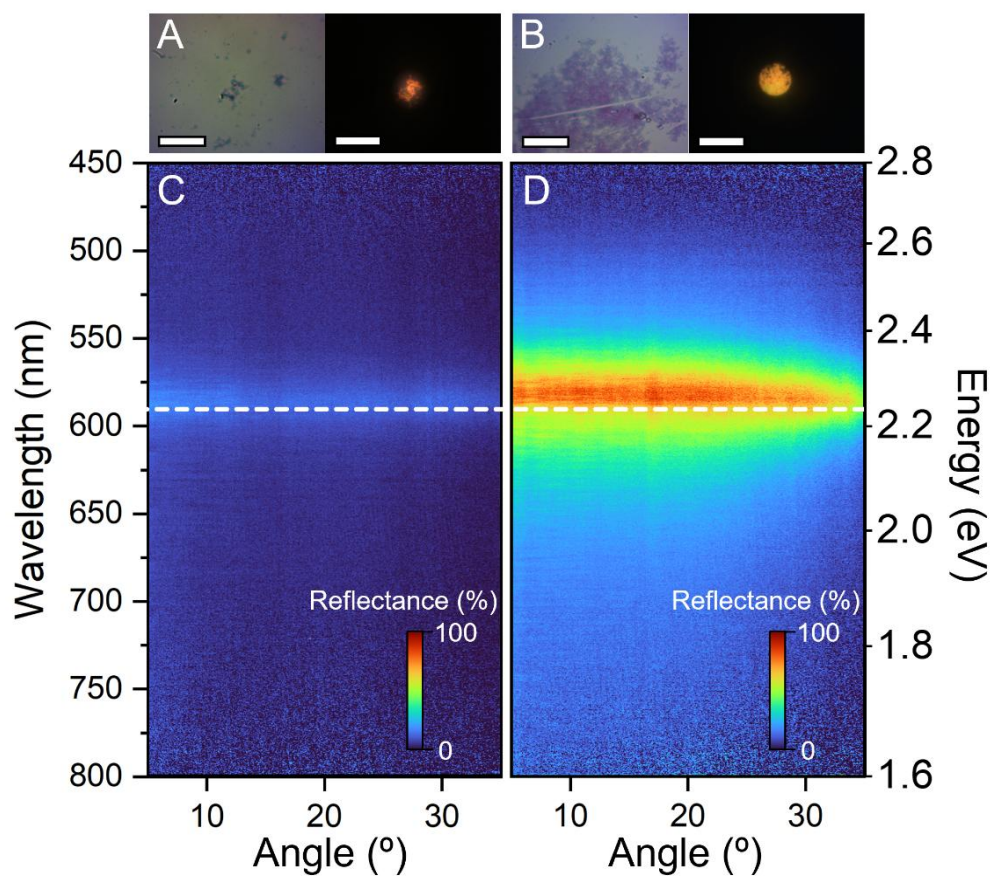

**Figure S8**. Single-flake optical analysis of flakes obtained with J-aggregate:PDDA molar ratio of 30. A-B) White-LED transmission (left panel) and halogen-lamp epi-illumination (right panel) microscopy images of two different flakes measured with size diameters of <15 μm (A, small) and >15 μm (B, large). The scale bar is 20 μm in all images. C-D) Angular reflectance for p-polarized light of single flakes with size diameters of <15 μm (C, small) and >15 μm (D, large). White dotted line represents the TDBC J-aggregate maximum of absorption at 587 nm.

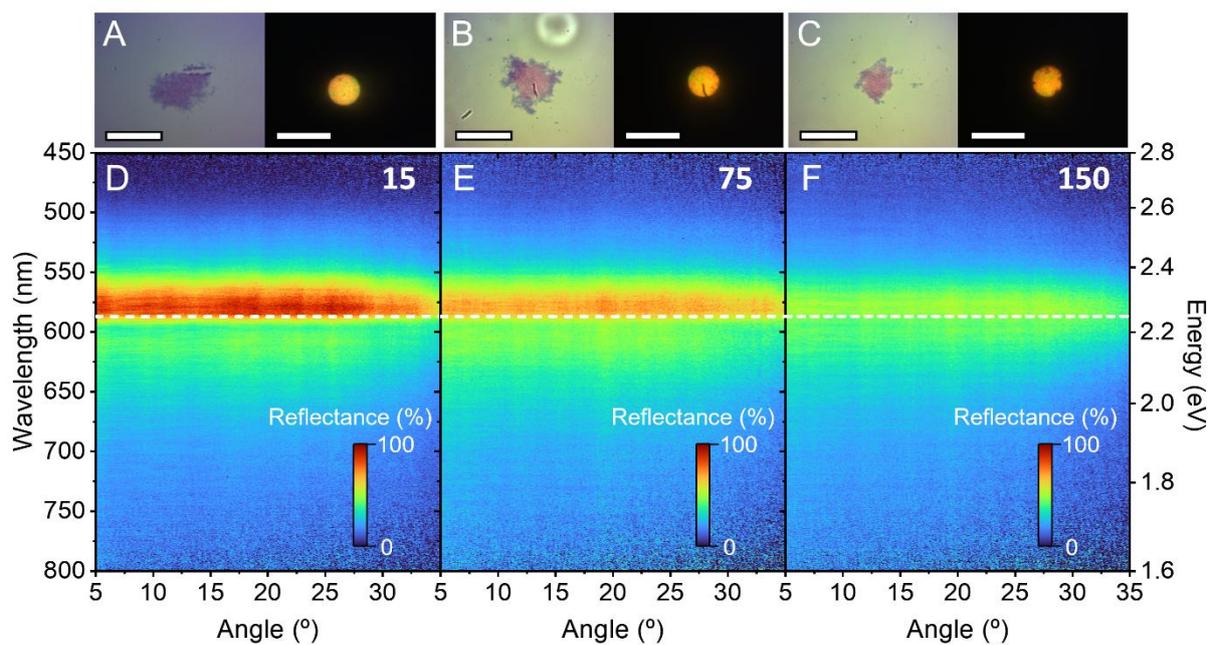

**Figure S9**. A-C) White-LED transmission (left panel) and halogen-lamp epi-illumination (right panel) microscopy images of single flakes with different J-aggregate:PDDA molar ratios, from left to right 15 (A), 75 (B) and 150 (C). D-F) Corresponding p-polarized light angular reflectance of single flakes prepared with different J-aggregate:PDDA molar ratios: 15 (D), 75 (E) and 150 (F). The scale bar is 20 μm in all images and the white dotted line represents the TDBC J-aggregate maximum of absorption at 587 nm.

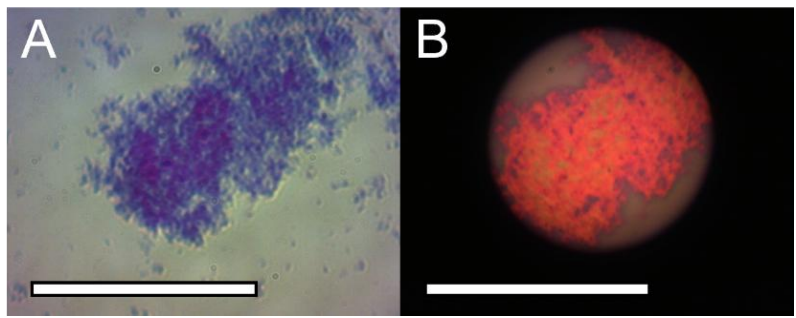

**Figure S10**. Flake with J-aggregate:PDDA molar ratios of 30, imaged using the same oil-immersion high NA lens used for FIS in the Kretschman Prism Coupling configuration. A-B) White-LED transmission (A) and halogen lamp epi-ilumination (B) microscopy images. The scale bar is 15 μm in both images. Note that the area of the spot in (B) determines the area in which the SEP is excited.